\documentclass{article}
\usepackage{spconf,amsmath,graphicx,hyperref}
\usepackage{latexsym}
\usepackage{booktabs}
\usepackage{multirow}
\usepackage{subfigure}
\usepackage{xcolor} 

\title{SRA: Semantic Relation-Aware Flowchart Question Answering}
\name{
Xinyu Li$^{1}$, Bowei Zou$^{2}$, Yuchong Chen$^{1}$, Yifan Fan$^{1}$, Yu Hong$^{1}$$^*$\thanks{$^*$ Corresponding author.}
}
\address{$^{1}$School of Computer Science and Technology, Soochow University, Suzhou, China\\
$^{2}$Institute for Infocomm Research, A*STAR, Singapore
}
\begin{document}
%
\maketitle
\begin{abstract}      
Flowchart Question Answering (FlowchartQA) is a multi-modal task that automatically answers questions conditioned on graphic flowcharts. 
Current studies convert flowcharts into interlanguages (e.g., Graphviz) for Question Answering (QA), which effectively bridge modal gaps between questions and flowcharts. More importantly, they reveal the link relations between nodes in the flowchart, facilitating a shallow relation reasoning during tracing answers. However, the existing interlanguages still lose sight of intricate semantic/logic relationships such as ``{\em Conditional}'' and ``{\em Causal}'' relations. This hinders the deep reasoning for complex questions. To address the issue, we propose a novel Semantic Relation-Aware (SRA) FlowchartQA approach. It leverages Large Language Model (LLM) to detect the discourse semantic relations between nodes, by which a link-based interlanguage is upgraded to the semantic relation based interlanguage. In addition, we conduct an interlanguage-controllable reasoning process. In this process, the question intention is analyzed with the aim to determine the depth of reasoning ({\em Shallow} or {\em Deep} reasoning), as well as the well-matched interlanguage. We experiment on the benchmark dataset FlowVQA. The test results show that SRA yields widespread improvements when upgrading different interlanguages like Graphviz, Mermaid and Plantuml.

\end{abstract}
\begin{keywords}
Multi-modality, FlowchartQA
\end{keywords}

\section{Introduction}
\begin{figure}[t]
  \centering
  \includegraphics[width=0.66\linewidth]{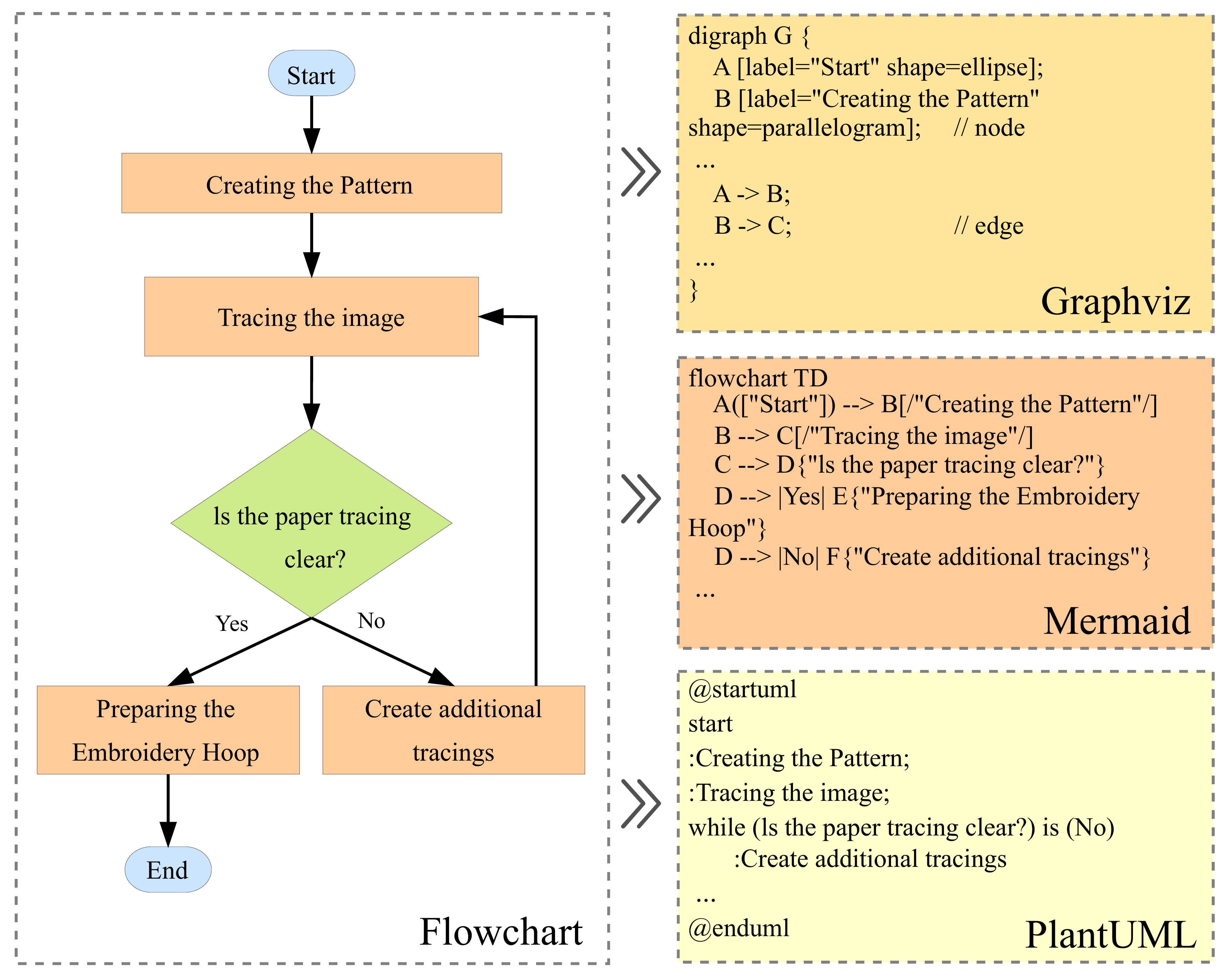} 
  \caption {Example of flowchart and interlanguages.}
  \label{fig:Figure 0}
\end{figure}

\begin{figure*}[t]
  \centering
  \includegraphics[width=0.8\linewidth]{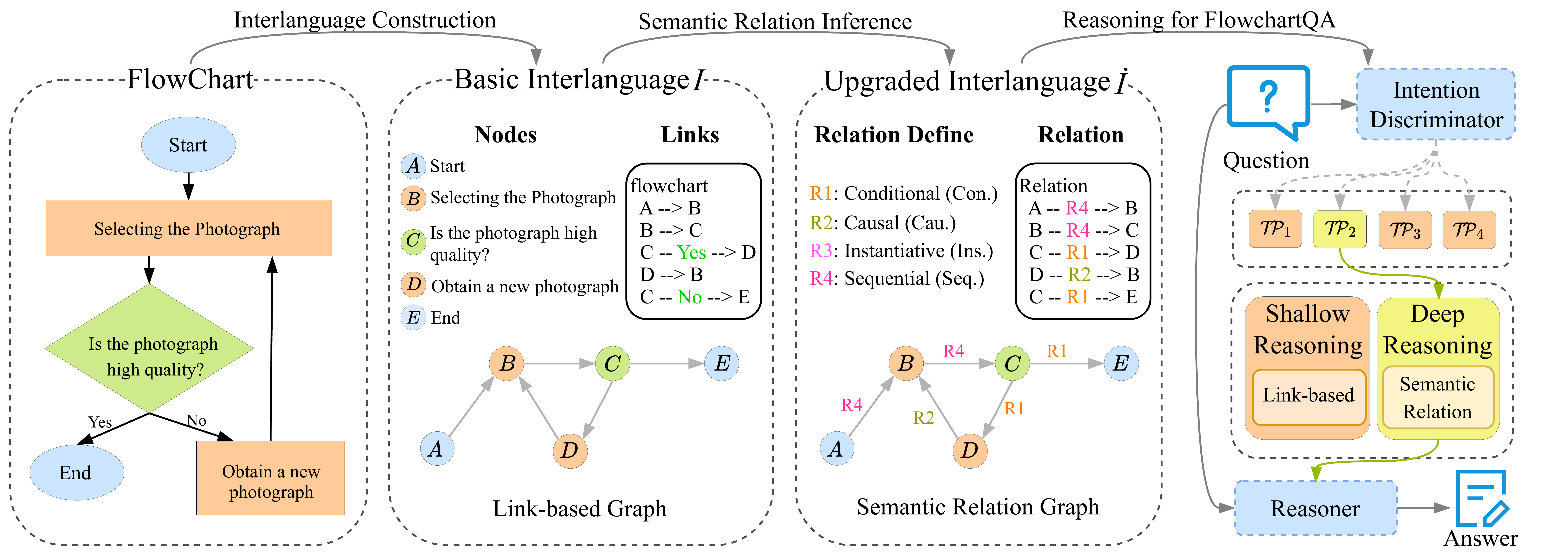} 
  \caption {SRA-based FlowcharQA architecture. $\mathcal{TP}_1$-$\mathcal{TP}_4$ denote different question types that correspond to distinct intentions.}
  \label{fig:Figure 1}
\end{figure*}

\label{sec:intro}
Flowchart is the multimodal data that combines diagrammatic structure with natural language. It serves to illustrate complex processes, workflows, and algorithms. Structurally, it consists of function nodes and edges, where a node contains the textual description, while a directed edge links the related nodes. Fig \ref{fig:Figure 0} provides an example of flowchart. FlowchartQA \cite{tannert-etal-2023-flowchartqa,singh-etal-2024-flowvqa,pan2024flowlearnevaluatinglargevisionlanguage,zhang2024multidimensionalevaluationflowchartcomprehension} aims to answer questions conditioned on flowcharts. It is a challenging task due to the requirement of cross-modal data analysis and semantics reasoning for QA.

The studies of FlowchartQA comprise three major trends, including 1) end-to-end cross-modal transformation \cite{singh-etal-2024-flowvqa,pan2024flowlearnevaluatinglargevisionlanguage,DBLP:conf/nips/LuQCXZZYLZ21IconQA}, 2) unstructured and structured data decoupling \cite{10.1007/978-981-97-5492-2_8GeFlowchartn,omasa2025arrowguidedvlmenhancingflowchart}, and 3) Vision Language Model (VLM) based flowchart understanding \cite{ye-etal-2025-beyond, wang2024qwen2vlenhancingvisionlanguagemodels,Llava,DBLP:conf/cvpr/HongWLXYJWWD0024CogAgent,DBLP:journals/corr/abs-2401-16420InternLM,DBLP:journals/corr/abs-2312-11805Gemini,claude,hurst2024gpt}. We concentrate on the third trend in this paper. In general, VLM-based approaches convert flowchart into a tailored semi-structured interlanguage, such as Mermaid\footnote{Mermaid: {\url{https://mermaid.js.org/}}}, Graphviz\footnote{Graphviz: {\url{https://graphviz.org/}}}, Plantuml\footnote{Plantuml: {\url{https://Plantuml.com/}}}. LLM is further prompted to understand the semantics of interlanguages and generate the answer accordingly. This category of approaches markedly improves the performance of FlowchartQA. 

However, in the VLM-based approaches, interlanguages merely depict the connection between nodes, and thus reveal a kind of stereotyped link relations--- relevance (as shown in Fig. \ref{fig:Figure 0}). In other words, they fail to embody the vivid semantic relations (e.g., causality and conditionality) between concrete contents of nodes. As a result, QA models can only perform a shallow reasoning over the data, instead of deep reasoning.

To address the issue, we propose SRA to enhance awareness of semantic relationships. It upgrades the existing interlanguages by replacing link relations with ideographic semantic relations. LLM is used to predict semantic relations for every pair of nodes in terms of their concrete contents. All relation types are summarized from the discourse relation taxonomy of the typical linguistic benchmark (Penn Discourse TreeBank 2.0 \cite{DBLP:conf/lrec/PrasadDLMRJW08}) instead of arbitrary selection. In addition, we design an interlanguage-controllable reasoning process. It enables the dynamic switching between shallow reasoning (on link relations) and deep reasoning (on semantic relations) for QA, in terms of question intentions.


We experiment on FlowVQA \cite{singh-etal-2024-flowvqa}. The test results demonstrate the practicability of SRA, which yields widespread improvements when different interlanguages are considered and different sizes of LLMs are used as backbones. At the best situation, it achieves an accuracy of about 71.3\%.

\section{Approach}
\label{sec:2_1}
\subsection{SRA-based FlowchartQA Architecture}
We enhance the baseline FlowchartQA model with SRA. The architecture is shown in Fig.\ \ref{fig:Figure 1}. The baseline \cite{ye-etal-2025-beyond} converts the flowchart into an interlanguage $\mathcal{I}$, and performs shallow reasoning $f_s(\ast)$ to predict the answer $\mathcal{A}$ for the question $Q$. It is formalized as $\mathcal{A}=f_s(\mathcal{I},\mathcal{Q})$. SRA upgrades the interlanguage $\mathcal{I}$ to the semantic relation-aware version $\mathcal{\dot{I}}$, and performs deep reasoning $f_d(\ast)$ to predict the answer. It is formalized as $\mathcal{A}=f_d(\mathcal{\dot{I}},\mathcal{Q})$. In SRA, interlanguage-controllable reasoning is allowed, where either shallow reasoning $f_s(\ast)$ or deep $f_d(\ast)$ is conducted given the question intention. Implementation details are as follows.

\begin{table}
\caption{Taxonomy of semantic relations.}
\centering
\small
\renewcommand{\arraystretch}{1} 
\resizebox{0.45\textwidth}{!}{
    \begin{tabular}{p{8cm}}
    \toprule[1pt]
    \textbf{Conditionality} (Definition): Node $O_B$ describes an outcome that depends on the condition presented in Node $O_A$. \\
    \ \ Example: \underline{$O_A$}-- {\em Finish your homework?}\ \  \underline{$O_B$}-- {\em Break}\ \ \underline{Relation} between $O_A$ and $O_B$-- Conditionality\\
    \midrule
    \textbf{Causality} (Definition): Node $O_A$ describes the Causal or action that directly causes the effect in Node $O_B$.\\
    \ \ Example: \underline{$O_A$}-- {\em Obtain a new photograph}\ \ \underline{$O_B$}-- {\em Selecting the Photograph}\ \ \underline{Relation} between $O_A$ and $O_B$-- Causality\\
    \midrule
    \textbf{Instantiation} (Definition): Node $O_B$ provides a specific instance or case of the general concept in Node $O_A$.\\
    \ \ Example: \underline{$O_A$}-- {\em Citrus Fruits}\ \ \underline{$O_B$}-- {\em Orange and Grapefruit}\ \ \underline{Relation} between $O_A$ and $O_B$-- Instantiation\\
    \midrule
    \textbf{Sequentiality} (Definition): Node $O_B$ describes an event that occurs chronologically after Node $O_A$. \\
    \ \ Example: \underline{$O_A$}-- {\em Mix the flour and water}\ \ \underline{$O_B$}-- {\em Knead the mixture into a dough}\ \ \underline{Relation} between $O_A$ and $O_B$-- Sequentiality\\
    \bottomrule[1pt]
    \end{tabular}
}
\label{tab:ablation}
\end{table}


\subsection{Basic Interlanguage and Shallow Reasoning}
\label{sec:2_2}
The baseline performs FlowchartQA on the link-based interlanguage. Specifically, we use TextFlow \cite{ye-etal-2025-beyond} to decompose the flowchart, dividing it into node pairs. During decomposition, VLM in TextFlow translates the visual image of flowchart to the analyzable structured data (e.g., Mermaid data \cite{ye-etal-2025-beyond}), which illustrates all nodes, edges, and labels in flowchart. The considered labels are limited to protogenic symbols ``{\em Yes}'' and ``{\em No}'' over edges (if have) as shown in Fig.\ \ref{fig:Figure 1}. We use such structured data as the basic interlanguage $\mathcal{I}$, which only reveals link relations and the simple logic of ``{\em AND-OR-NOT}''. In experiments, different formats of structured data are considered for evaluation, including Mermaid and Graphviz \cite{ye-etal-2025-beyond}.

The LLM Qwen-2.5 \cite{qwen2025qwen25technicalreport} is used to infer the answer $\mathcal{A}$, where question $\mathcal{Q}$ and interlanguage $\mathcal{I}$ are packed in a query as input. Qwen-2.5 runs in the zero-shot setting, and there is no explicit instruction provided to induce a task-specific Chain-of-Thought (CoT). Consequently, Qwen-2.5 performs the relation reasoning canonically, upon the shallow node-wise relation knowledge (i.e., links and simple logic).

\subsection{Semantic Relation-aware Interlanguage}
\label{sec:2_3}
\subsubsection{Relation Taxonomy}
\label{sec:2_3_1}
We construct a tailored semantic relation taxonomy for interlanguage upgrading. This taxonomy comprises four relation types, including {\em conditional}, {\em causal}, {\em instantiated} and {\em sequential}. Table \ref{tab:ablation} provides their definitions and examples. All of them are derived from the completed discourse relation taxonomy of Penn Discourse TreeBank 2.0 (PDTB v2) \cite{DBLP:conf/lrec/PrasadDLMRJW08}, which are originally used to automatically analyze semantic relationships among sentence-level texts.

Note that PDTB actually contains 4 main categories of 11 sense-level relation types. Most of them (except the four types mentioned above) are disabled in this study. It is because 1) the disabled relations are incompatible with interactive logic between flowchart nodes, and 2) thus they will occur as noises in the upgraded interlanguage. Deepseek-R1 \cite{DBLP:journals/corr/abs-2501-12948deepseekr1} is used to verify compatibility. There are 1K node pairs used for verification, which are randomly selected from the training set.

\subsubsection{Semantic Relation Recognition for Upgrading}
\label{sec:2_3_2}
Our SRA constructs a semantic relation-aware interlanguage $\mathcal{\dot{I}}$ in terms of the above relation taxonomy. It is implemented by upgrading the basic interlanguage $\mathcal{I}$, where the link relation tag of each pair of nodes is replaced with their semantic relation tag. Relation recognition is used for tag relabeling. 

Specifically, for each pair of nodes in $\mathcal{I}$, we use Deepseek-R1 \cite{DBLP:journals/corr/abs-2501-12948deepseekr1} to recognize their semantic relation. Tag relabeling is performed synchronously. During relation recognition, a task-specific instruction is used to induce a two-phase Chain-of-Thought (CoT) of Deepseek-R1. First, it is driven to read the contents of nodes, and analyze their semantic relation accordingly. Second, by the analysis result, it determines the most well-matched relation tag in the taxonomy. All the relation definitions (Table \ref{tab:ablation}) are disclosed in the input query, along with the basic interlanguage $\mathcal{I}$. The output is limited to the triple $\mathcal{T}$ of nodes {\em d} and relation {\em r}, i.e., $\mathcal{T}_{ij}$=\{$n_i$, $r_{ij}$, $n_j$\}.

\subsubsection{Deep Reasoning}
\label{sec:2_3_3}
Using the upgraded language $\mathcal{\dot{I}}$, Qwen-2.5 is prompted to execute deep reasoning, where deeper knowledge is provided in the query. Specifically, relation-centered triples $\mathcal{T}$ are given to exhibit the diversified correlations, while the taxonomy interprets concrete senses of relations. Similar to shallow reasoning, zero-shot setting is used during deep reasoning, and CoT is also omitted. Though, strict constraints are used, which instruct Qwen-2.5 to sufficiently perceive triples $\mathcal{T}$, taxonomy and question before generating an answer.

\subsection{Interlanguage-controllable Reasoning}
\label{sec:2_4}
FlowchartQA tackles different types of questions. Some of them are straight, with less or no intention to resolve complicated relations. For example, in the four types of questions ($\mathcal{TP}_{1-4}$) in the benchmark FlowVQA\cite{singh-etal-2024-flowvqa}, only the type of ``{\em Applied Scenario}'' ($\mathcal{TP}_{2}$) heavily relies on deep relation analysis (e.g., condition or sequentiality of actions). By contrast, the type of ``{\em Topology}'' ($\mathcal{TP}_{4}$) only concerns structural information (e.g., see the question of ``{\em how many nodes in the chart?}''), which has nothing to do with relation reasoning.

Applying deep reasoning (Section \ref{sec:2_3_3}) on those straight questions causes unnecessary consumption of computing resources and time. Therefore, we develop an interlanguage-controllable reasoning method. It uses a LLM-based discriminator to determine the question type, outputting a ``{\em Straight}'' or ``{\em Complicated}'' question tag. For straight questions, it invokes LLM to perform shallow reasoning using the basic interlanguage $\mathcal{I}$ (Section \ref{sec:2_2}). For complicated cases, deep reasoning is performed using the upgraded interlanguage $\mathcal{\dot{I}}$ (Section \ref{sec:2_3_3}). During implementation, LLaMA-3 (8B) \cite{Llama} is used as the discriminator. This discriminator is fine-tuned on 2K randomly-selected instances from FlowVQA training set.

\setlength{\tabcolsep}{4pt}
\begin{table}[t]
\caption{Comparison results ({\em Acc. \%}) on the test set.}
\small
\centering
\renewcommand{\arraystretch}{0.9} 
\resizebox{1\columnwidth}{!}{
\begin{tabular}{c|c|c|cccc}
\toprule[1pt]
\multirow{2}{*}{\textbf{Backbone}} & \multirow{2}{*}{\textbf{Method}} & \multirow{2}{*}{\textbf{Overall}} & \multicolumn{4}{c}{\textbf{Separate Test Sets}} \\
\cmidrule(lr){4-7} &  &  & \textbf{$\mathcal{TP}_{1}$} & \textbf{$\mathcal{TP}_{2}$} & \textbf{$\mathcal{TP}_{3}$} & \textbf{$\mathcal{TP}_{4}$}  \\

\midrule

\multirow{1}{*}{\textbf{GPT-4o}}
& \multicolumn{1}{l|}{Image} & 63.7 & 69.5 & 65.6 & 68.8 & 57.9 \\
\midrule

\multirow{6}{*}{\textbf{Qwen-2.5 (72B)}} 
& \multicolumn{1}{l|}{Graphviz} & 68.7 & 79.8 & 63.6 & 72.5 & 64.2  \\
& \ \ \ \ + SRA & \textbf{71.3} & 83.5 & 73.1 & 72.0 & 64.4  \\
\cmidrule{2-7}
& \multicolumn{1}{l|}{Mermaid} & 69.5 & 83.8 & 68.1 & 73.1 & 62.0  \\
& \ \ \ \ + SRA & 70.1 & 84.5 & 72.9 & 71.7 & 61.2  \\
\cmidrule{2-7}
& \multicolumn{1}{l|}{PlantUML} & 57.3 & 79.8 & 63.6 & 70.2 & 38.1  \\
& \ \ \ \ + SRA & 60.0 & 82.0 & 72.9 & 68.8 & 39.6  \\
\midrule



\multirow{6}{*}{\textbf{Qwen-2.5 (7B)}} 
& \multicolumn{1}{l|}{Graphviz} & 59.4 & 78.8 & 59.3 & 54.6 & 52.0  \\
& \ \ \ \ + SRA & 61.0 & 81.3 & 65.1 & 54.3 & 52.0  \\
\cmidrule{2-7}
& \multicolumn{1}{l|}{Mermaid} & 60.6 & 80.0 & 60.3 & 56.4 & 53.1  \\
& \ \ \ \ + SRA & \textbf{62.0} & 80.8 & 66.1 & 56.7 & 53.2  \\
\cmidrule{2-7}
& \multicolumn{1}{l|}{PlantUML} & 53.0 & 82.8 & 57.8 & 57.8 & 34.4  \\
& \ \ \ \ + SRA & 54.7 & 83.3 & 65.6 & 57.5 & 34.5  \\
\bottomrule[1pt]

\end{tabular}
}
\label{tab:mainex}
\end{table}

\section{Experimentation}
\subsection{Dataset and Evaluation}
We experiment on FlowVQA \cite{singh-etal-2024-flowvqa}. There are 12,938 instances can be used for training in FlowVQA, while 9,475 for test. For comparison purposes, we follow the practice of previous work \cite{ye-etal-2025-beyond} to form the test set using 1,978 instances, which are QA pairs of 197 flowcharts. Accuracy ({\em Acc}.) is used as the evaluation metric. In addition, GPT-4o is used to determine positive and negative examples by comparing predictions with ground-truth data \cite{singh-etal-2024-flowvqa,pan2024flowlearnevaluatinglargevisionlanguage,ye-etal-2025-beyond}. This is recognized as a semantic matching method for correctness verification.


\subsection{Main Results}
We compare SRA to image-based FlowchartQA \cite{ye-etal-2025-beyond}, where GPT-4o is used for image2text. We also compare with three interlanguage-based baselines. They uniformly use Qwen-2.5 (7B/72B) as backbones, though infer answers on different interlanguages, including {\em Graphviz}, {\em Mermaid} \cite{mermaid} and {\em Plantuml} \cite{DBLP:graphviz_plantuml} . Note Graphviz and Mermaid serve as link-based structured data (Section \ref{sec:2_2}). Plantuml is a semi-structured language, manifesting as a program holding execution logic. 

Table \ref{tab:mainex} shows the comparison results, including the comprehensive performance ({\em Acc}.) on the \underline{\textbf{overall}} test set, as well as local performance on the separate subsets of different question types. The types comprise {\em Fact Retrieval} ($\mathcal{TP}_1$), {\em Applied Scenarios} ($\mathcal{TP}_2$), {\em Flow Reference} ($\mathcal{TP}_3$) and {\em Topology} ($\mathcal{TP}_4$). It can be observed from Table \ref{tab:mainex} that SRA obtains higher comprehensive performance compared to baselines no matter what backbones are used. The additional finding is that SRA yields the most substantial improvements for interpretability-centered questions ($\mathcal{TP}_2$). The improvement is up to 9.5\% at best (compared to {\em Graphviz}). Though, SRA achieves comparable performance for other three types.

\begin{figure}[t]
  \includegraphics[width=0.8\columnwidth]{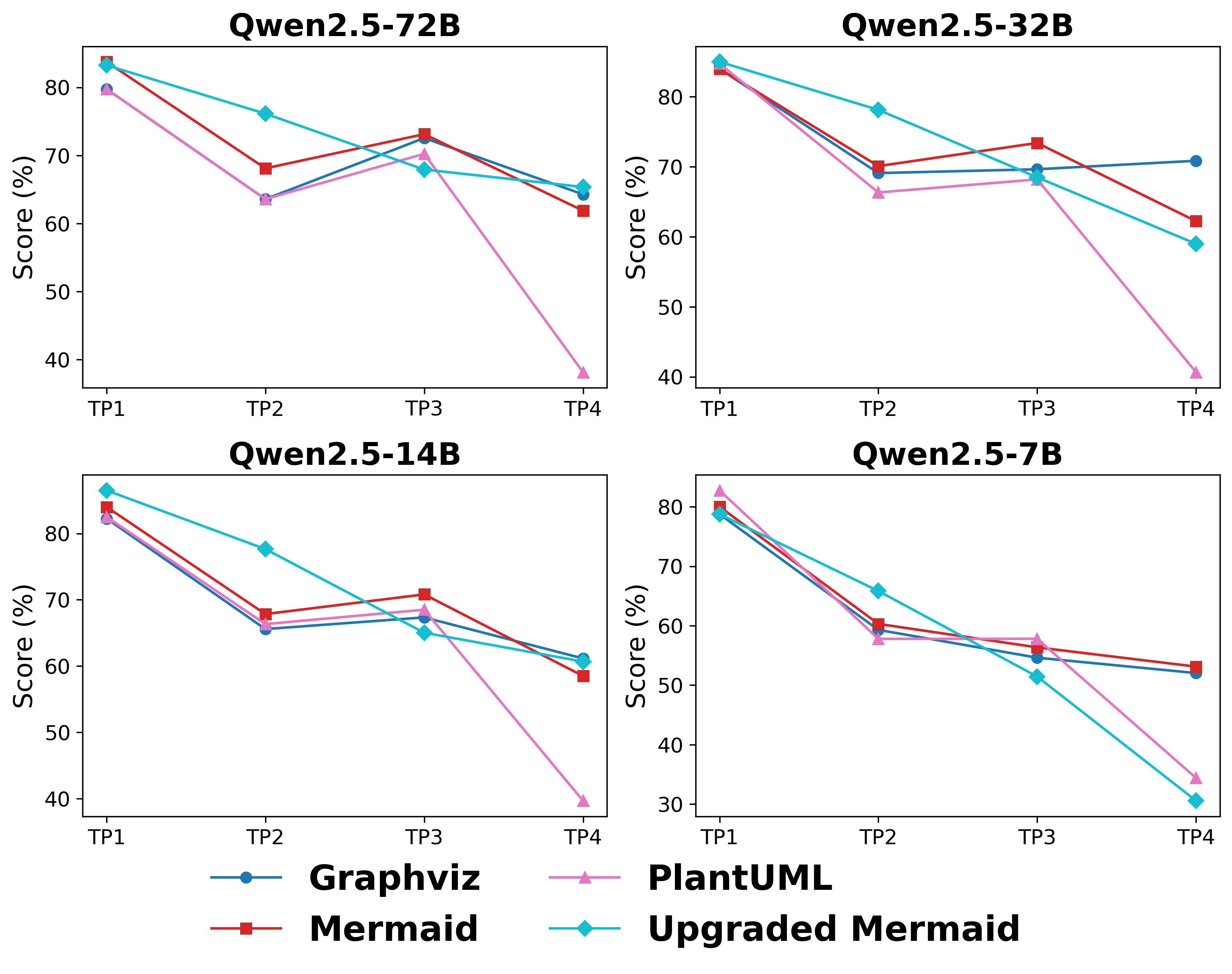}
  \centering
  \caption{Multi-interlanguage mono-directional verification.}
  \label{fig:ablation}
\end{figure}

\subsection{Necessity of Interlanguage-controllable Setting}
The local performance of SRA (in Table \ref{tab:mainex}) for the above four question types ($\mathcal{TP}_{1-4}$) seemingly proves the necessity of interlanguage-controllable reasoning. Nevertheless, the verification is indirect because the control is actuated by the automatic discriminator. This raises our inspection about that pseudo-positive samples may cause good fortune.

To capture direct evidence, we disable the interlanguage-controllable reasoning module. On this basis, FlowchartQA is independently carried out for a single type of questions, using different interlanguages (Graphviz, Mermaid, Plantuml and upgraded Mermaid). This enables multi-interlanguage mono-directional verification.  Fig.\ref{fig:ablation} shows the verification result for every question type, where different versions of Qwen-2.5 are used. It can be observed that, for $\mathcal{TP}_2$, the adoption of the upgraded interlanguage always leads to a much higher performance (blue line in Fig.\ \ref{fig:ablation}). On the contrary, comparable or unstable performance is obtained for other tree question types, which are specified as straight questions in Section \ref{sec:2_4}. The verification results directly prove the necessity of interlanguage-controllable setting.

\subsection{Error Analysis for Discriminator}
Question type discrimination in SRA serves as an intermediate operation, which is conducted before the interlanguage-controllable reasoning process. Thus, the erroneous question types it produces will mistakenly lead the subsequent shallow or deep reasoning. Actually, a separate experiment shows that the discriminator obtains an accuracy of 85.8\% on the test set. The negative effects of 14.2\% errors cannot be overlooked.

\begin{figure}[t]
\centering
\begin{subfigure}[Number of errors]{
    \includegraphics[width=0.2\textwidth]{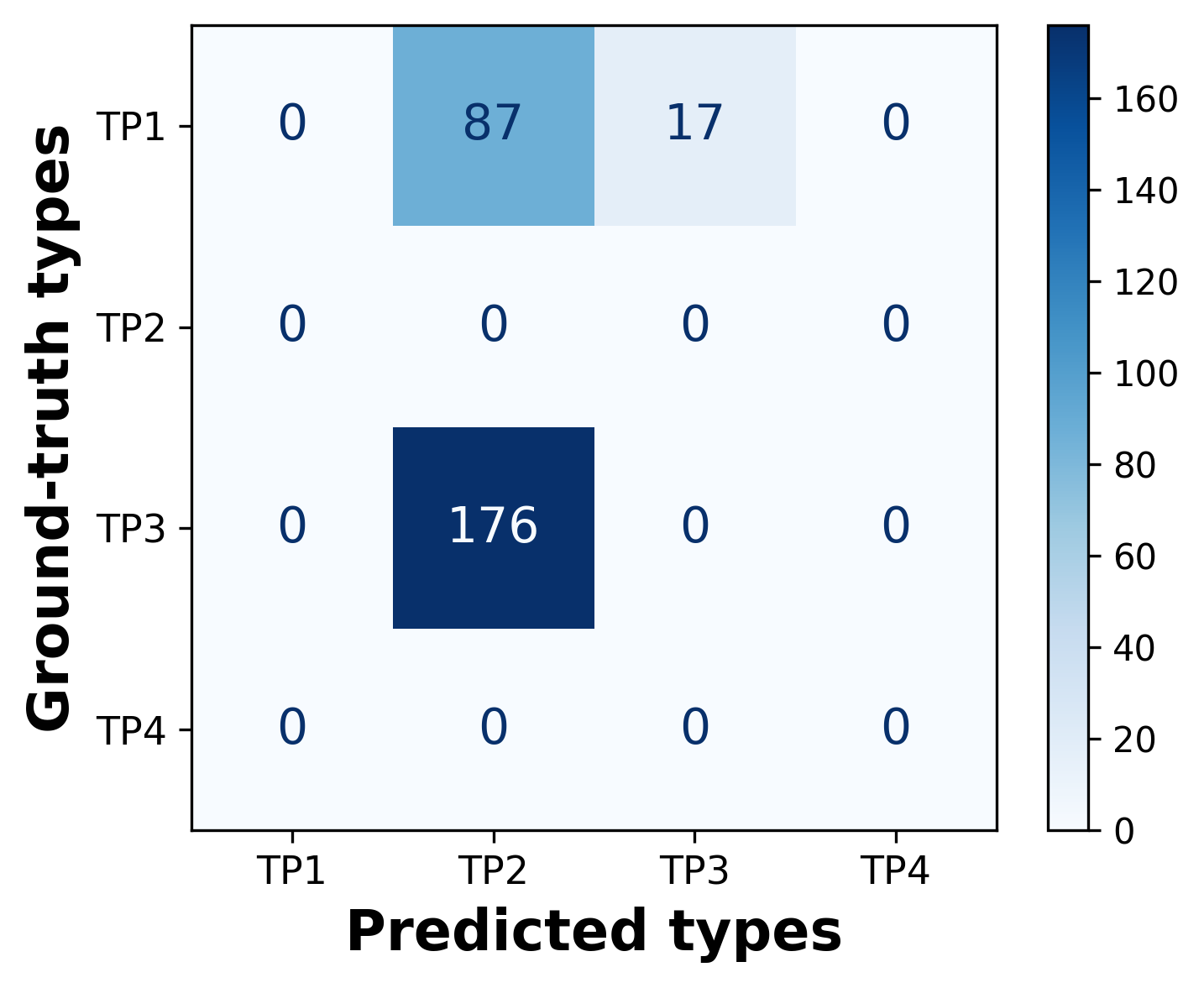}
    \label{fig:1}
}
\end{subfigure}
\quad
\begin{subfigure}[Error analysis]{
    \includegraphics[width=0.2\textwidth]{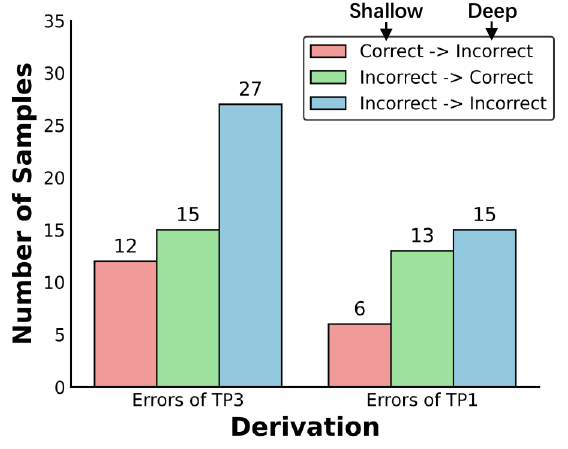}
    
    \label{fig:2}
}
\end{subfigure}
\caption{Error analysis (Qwen-2.5-72B and Graphviz are used).}\label{fig:other_exper}
\end{figure}

To gain a clear insight into the effects, we investigate the distribution of errors across the four types of questions. All the instances in the test set are considered for investigation. Fig.\ \ref{fig:1} shows the error distribution, where each heat block indicates the number of errors. It reveals that the main errors derive from $\mathcal{TP}_1$ and $\mathcal{TP}_3$ questions, some of which are mistakenly determined as $\mathcal{TP}_2$. Considering that $\mathcal{TP}_2$ is dealt with by deep reasoning in the controllable process, we further verify the accuracy of deep reasoning on these misjudged questions, and compare it with shallow reasoning. The subgraph (b) in Fig.\ \ref{fig:2} shows the comparison results. It illustrates that some of the misjudged questions are indeed correctly answered by shallow reasoning but incorrectly by deep (red histogram). By contrast, the inverse effect is relatively more obvious (green histogram). This confirms the subsistent fault-tolerant ability of deep reasoning.


\subsection{Effects of LLMs for Reasoning}
LLM-based semantic relation generation is the core of interlanguage upgrading. We verify the reliability of two LLMs, including DeepSeek-R1 and GPT-4o. Manual inspection is conducted on 500 randomly-selected samples. It reveals that Deepseek-R1 outperforms GPT-4o (86.2\% vs 80.0\% for {\em Acc}).




\section{Conclusion}
To enhance FlowchartQA, we proposed upgrading the structured interlanguage using semantic relations, with the aim of increasing the depth of reasoning. In addition, we develop an interlanguage-controllable reasoning method, which selectively invokes shallow and deep reasoning towards different question types. Experiments demonstrate the effectiveness of our approach. In the future, we will study the self-contained knowledge-augmented approach, detecting the invisible knowledge in the graph for comprehensive understanding, such as topics and multi-hop implicit relations. 


\vfill\pagebreak
\bibliographystyle{IEEEbib}
\bibliography{refs_new}

\end{document}